\documentstyle[12pt,axodraw]{article}

\catcode`@=12
\begin{document}
\thispagestyle{empty}
\begin{flushright} 
UCRHEP-T309\\ 
July 2001\
\end{flushright}
\vspace{0.5in}
\begin{center}
{\LARGE	\bf Neutrino, Lepton, and Quark Masses\\ in Supersymmetry\\}
\vspace{1.5in}
{\bf Ernest Ma\\}
\vspace{0.2in}
{\sl Physics Department, University of California, Riverside, 
California 92521\\}
\vspace{1.5in}
\end{center}
\begin{abstract}\
The recently proposed model of neutrino mass with no new physics beyond the 
TeV energy scale is shown to admit a natural and realistic supersymmetric 
realization, when combined with another recently proposed model of quark 
masses in the context of a softly broken U(1) symmetry.  Four Higgs doublets 
are required, but two must have masses at the TeV scale.  New characteristic 
experimental predictions of this synthesis are discussed.
\end{abstract}
\newpage
\baselineskip 24pt

In the minimal standard model of fundamental particle interactions, neutrinos 
are massless.  In the minimal supersymmetric standard model (MSSM), they are 
still massless, because of the imposition of additive lepton-number 
conservation.  Although the assignment of lepton number(s) is by no means 
unique \cite{mang}, a minimal scenario for neutrino mass is to assume the 
conservation of a discrete $Z_2$ (odd-even) symmetry which is odd for all 
leptons and even for all others.  By the addition of three neutral 
singlet lepton superfields $N_i$ with allowed large Majorana masses, the 
usual doublet neutrinos $\nu_i$ will then obtain small masses through the 
famous seesaw mechanism \cite{seesaw}. 

The conventional wisdom is that $m_N$ must be very large, say of order 
10$^{13}$ GeV or greater, for $m_\nu$ to be much less than 1 eV.  However, 
it has been shown recently \cite{ma00} that $m_N \sim 1$ TeV is possible 
(and natural) if there exists a second Higgs doublet with $m^2 > 0$ so 
that its vacuum expectation value (VEV) is naturally small, say of order 
1 MeV.  This is achieved by an appropriate assignment of additive lepton 
number which is softly broken in the scalar sector.  More recently, a 
model of quark masses is proposed \cite{ma01}, where the smallness 
of $m_u, m_d, m_s$ compared to $m_c, m_b, m_t$ and the pattern of the 
charged-current mixing matrix may be understood in a similar way. 
In this paper the two proposals are shown to be naturally combined in a 
supersymmetric model with four Higgs doublets, in the context of a \underline 
{single} softly broken U(1) symmetry.

The gauge group is the standard one, i.e. $SU(3)_C \times SU(2)_L \times 
U(1)_Y$.  The particle content is the usual three families of quark and 
lepton superfields, with the addition of three neutral singlet superfields 
$N_i$ and four (instead of two) Higgs superfields.  Each matter superfield 
(all defined to be left-handed) transforms under an assumed global U(1) 
symmetry as follows:
\begin{eqnarray}
0 &:& (t,b), t^c, b^c, s^c, d^c, N_i, (h_1^0,h_1^-), (h_2^+,h_2^0) \\ 
1 &:& (\nu_i,l_i), c^c, (h_3^0,h_3^-) \\ 
-1 &:& (c,s), (u,d), \tau^c, (h_4^+,h_4^0) \\ 
2 &:& u^c \\ 
-2 &:& \mu^c, e^c
\end{eqnarray}
Let $h^0_{1,2,3,4}$ acquire VEVs equal to $v_{1,2,3,4}$ respectively, then 
the quark mass matrices are given by \cite{ma01}
\begin{equation}
{\cal M}_u = \left[ \begin{array} {c@{\quad}c@{\quad}c} f_u v_4 & 0 & 0 \\ 
f_{cu} v_4 & f_c v_2 & 0 \\ 0 & f_{tc} v_4 & f_t v_2 \end{array} \right], ~~~ 
{\cal M}_d = \left[ \begin{array} {c@{\quad}c@{\quad}c} f_d v_3 & f_{ds} v_3 
& f_{db} v_3 \\ 0 & f_s v_3 & f_{sb} v_3 \\ 0 & 0 & f_b v_1 \end{array} 
\right],
\end{equation}
where the freedom to rotate among $(c,s)$ and $(u,d)$ has been used to set the 
$uc^c$ element to zero and the freedom to rotate among $(b^c,s^c,d^c)$ has 
been used to set the 3 lower off-diagonal entries of ${\cal M}_d$ to zero. 
Similarly, the charged-lepton mass matrix is given by
\begin{equation}
{\cal M}_l = \left[ \begin{array} {c@{\quad}c@{\quad}c} f_e v_3 & 0 & 0 \\ 
0 & f_\mu v_3 & 0 \\ f_{\tau e} v_3 & f_{\tau \mu} v_3 & 
f_\tau v_1 \end{array} \right],
\end{equation}
whereas the neutrino mass matrix linking $\nu_i$ to $N_j$ is proportional 
to $v_4$, but otherwise arbitrary.

If the assumed U(1) symmetry is unbroken, then $v_3 = v_4 = 0$.  This means 
that $m_u = m_d = m_s = 0$ and $m_e = m_\mu = m_{\nu_i} = 0$, i.e. only 
$t, b, c,$ and $\tau$ are massive. [Of course $N_j$ have allowed large 
Majorana masses, but there would be no Dirac mass matrix linking them to 
$\nu_i$.] To see how $v_3$ and $v_4$ become nonzero but small, consider 
the Higgs sector of this model.  The terms $H_1 H_2$ and $H_3 H_4$ are 
allowed by U(1) invariance, thus guaranteeing that appropriately large 
higgsino masses are present in the $6 \times 6$ (instead of the usual 
$4 \times 4$) neutralino mass matrix.   The terms $H_1 H_4$ and $H_2 H_3$ 
break U(1) softly, thus it is natural for their coefficients to be small 
\cite{thooft}, which allow $v_4 << v_1$ if $m_4^2 > 0$ while $m_1^2 < 0$ 
and $v_3 << v_2$ if $m_3^2 > 0$ while $m_2^2 < 0$, as explained in 
Refs.~[3,4]. [The $L_i H_{2,4}$ terms are forbidden by the unbroken $Z_2$ 
lepton parity discussed earlier.]

Since $m_t = f_t v_2$ and $m_b = f_b v_1$, the natural magnitude of $v_2$ 
is $10^2$ GeV and that of $v_1$ is a few GeV.  Hence it is natural as well 
for $v_3 \sim 10^2$ MeV and $v_4 \sim$ a few MeV.  A glance at Eqs.~(6) and 
(7) shows that these are indeed very realistic values.  Since $m_\nu \simeq 
f^2 v_4^2/m_N$, this also means that $m_N \sim$ a few TeV is realistic, as 
shown in Ref.~[3].  Note that Eqs.~(29), (31), (32), (33), and (35) of 
Ref.~[4] are unchanged (except of course $m_2$ and $v_2$ there are redefined 
as $m_3$ and $v_3$ here) because $f_b v_1 = m_b$ even though $v_1$ here is 
numerically much smaller.  Hence the constraints due to flavor-changing 
neutral currents (FCNC) in the $down$ sector are all satisfied provided that
\begin{equation}
m_3 > 3.23 \left( {0.3~{\rm GeV} \over v_3} \right) ~{\rm TeV},
\end{equation}
i.e. Eq.~(30) of Ref.~[4]. In the case of $D^0 - \overline {D}^0$ mixing, 
Eq.~(34) of Ref.~[4] becomes
\begin{equation}
{\Delta m_{D^0} \over m_{D^0}} \simeq {B_D f_D^2 v_2^2 \over 3 m_4^2} f_c^2 
f_{cu}^2 {m_u \over m_c^3} < 2.5 \times 10^{-14}.
\end{equation}
Using $f_D = 150$ MeV, $B_D = 0.8$, $f_c v_2 = m_c = 1.25$ GeV, and $m_u = 4$ 
MeV, this implies
\begin{equation}
m_4 > 2.77 \left( {f_{cu} \over 0.1} \right) {\rm TeV}.
\end{equation}

The Higgs potential of this model is given by
\begin{eqnarray}
V &=& \sum_i m_i^2 H_i^\dagger H_i + [m_{12}^2 H_1 H_2 + m_{34}^2 H_3 H_4 + 
m_{14}^2 H_1 H_4 + m_{23}^2 H_2 H_3 + h.c.] \nonumber \\ &+& {1 \over 2} 
g_1^2 \left[ -{1 \over 2} H_1^\dagger H_1 + {1 \over 2} H_2^\dagger H_2 - 
{1 \over 2} H_3^\dagger H_3 + {1 \over 2} H_4^\dagger H_4 \right]^2 + 
{1 \over 2} g_2^2 \sum_\alpha |\sum_i H_i^\dagger \tau_\alpha H_i|^2,
\end{eqnarray}
where $\tau_\alpha (\alpha = 1,2,3)$ are the usual SU(2) representation 
matrices.  Let $\langle h_i^0 \rangle = v_i$, then the minimum of $V$ is
\begin{eqnarray}
V_{min} &=& \sum_i m_i^2 v_i^2 + 2 m_{12}^2 v_1 v_2 + 2 m_{34}^2 v_3 v_4 + 
2 m_{14}^2 v_1 v_4 + 2 m_{23}^2 v_2 v_3 \nonumber \\ &+& {1 \over 8} 
(g_1^2 + g_2^2) (v_1^2 - v_2^2 + v_3^2 - v_4^2)^2,
\end{eqnarray}
where all parameters have been assumed real for simplicity.  The 4 equations 
of constraint are
\begin{eqnarray}
0 &=& m_1^2 v_1 + m_{12}^2 v_2 + m_{14}^2 v_4 + {1 \over 4} (g_1^2 + g_2^2) 
v_1 (v_1^2 - v_2^2 + v_3^2 - v_4^2), \\ 
0 &=& m_2^2 v_2 + m_{12}^2 v_1 + m_{23}^2 v_3 - {1 \over 4} (g_1^2 + g_2^2) 
v_2 (v_1^2 - v_2^2 + v_3^2 - v_4^2), \\ 
0 &=& m_3^2 v_3 + m_{34}^2 v_4 + m_{23}^2 v_2 + {1 \over 4} (g_1^2 + g_2^2) 
v_3 (v_1^2 - v_2^2 + v_3^2 - v_4^2), \\ 
0 &=& m_4^2 v_4 + m_{34}^2 v_3 + m_{14}^2 v_1 - {1 \over 4} (g_1^2 + g_2^2) 
v_4 (v_1^2 - v_2^2 + v_3^2 - v_4^2).
\end{eqnarray}
A solution with $v_4 << v_3 << v_1 << v_2$ is then possible with the result
\begin{equation}
v_2 \simeq {-m_2^2 \over {1 \over 4} (g_1^2 + g_2^2)}, ~~~ v_1 \simeq 
{-m_{12}^2 v_2 \over {m_1^2 + m_2^2}},
\end{equation}
and
\begin{equation}
v_3 \simeq {-m_{23}^2 v_2 \over m_3^2 - {1 \over 4} (g_1^2 + g_2^2) v_2^2}, ~~~v_4 \simeq {-m_{14}^2 v_1 - m_{34}^2 v_3 \over m_4^2 + {1 \over 4} (g_1^2 + 
g_2^2) v_2^2}.
\end{equation}
The $H_{1,2}$ doublets are essentially those of the MSSM, while $H_3$ and 
$H_4$ have masses $m_3$ and $m_4$ respectively at the TeV scale, as 
constrained phenomenologically by Eqs.~(8) and (10).  Once produced, 
the dominant decays of $H_{1,2}$ are the same as in the MSSM, i.e. into 
$t, b, c$ and $\tau$ states.  Their decay branching fractions into light 
fermions depend on $H_1 H_4$ and $H_2 H_3$ mixing, but since they are 
very much suppressed, it will be difficult to distinguish them from those 
of the MSSM.  If $H_3$ and $H_4$ are produced, then their decays will be 
the decisive evidence of this model.  As discussed in Ref.~[3], the decays
\begin{equation}
h_4^+ \to l_i^+ N_j, ~~~{\rm then} ~N_j \to l_k^\pm W^\mp,
\end{equation}
will determine the relative magnitude of each element of the neutrino mass 
matrix.  The difference in the present model is that $H_4$ also couples to 
$(u,d) u^c$, $(c,s) u^c$, and $(t,b) c^c$.  This means that the three-body  
decay of $N$ is actually dominant \cite{mrw}, i.e.
\begin{equation}
N \to \nu (l) + 2 ~{\rm quark~jets}.
\end{equation}
Of course, this still carries the relevant information on the 
neutrino mass matrix by the flavor of the charged lepton in the 
final state.

In the model of Ref.~[4], lepton flavor is assumed conserved, but it cannot 
be maintained in the presence of neutrino oscillations.  Here $H_3$ couples 
to both quarks and leptons together with $H_1$ according to ${\cal M}_l$ of 
Eq.~(7).  Following the discussion given in Ref.~[4], the FCNC 
effects in the charged-lepton sector are thus contained in the term
\begin{equation}
f_\tau \bar \tau_L \tau_R \left[ \bar h_1^0 - {v_1 \over v_3} \bar h_3^0 
\right] + h.c.,
\end{equation}
where $\tau_{L,R}$ are not mass eigenstates and have to be rotated using 
Eq.~(7).  The analog of Eq.~(28) of Ref.~[4] is then
\begin{eqnarray}
&& \left[ {v_3 \over v_1} \bar h_1^0 - \bar h_3^0 \right] \left[ f_{\tau \mu} 
(\bar \tau_L \mu_R + {m_\mu \over m_\tau} \bar \mu_L \tau_R) + f_{\tau e} 
(\bar \tau_L e_R + {m_e \over m_\tau} \bar e_L \tau_R) \right. \nonumber \\ 
&& \left. + {f_{\tau \mu} f_{\tau e} v_3 \over m_\tau^2} (m_\mu \bar \mu_L 
e_R + m_e \bar e_L \mu_R) \right] + h.c.
\end{eqnarray}
The most stringent bounds on $f_{\tau \mu}$ and $f_{\tau e}$ come from 
$\tau \to \mu \mu \mu$ and $\tau \to e \mu \mu$ through $h_3^0$ exchange.
Using $m_3 = 3.23$ TeV, $v_3 = 0.3$ GeV, and $f_{\tau e} = 1$, the 
fraction
\begin{equation}
{\Gamma (\tau \to e \mu \mu) \over \Gamma (\tau \to \nu_\tau e \nu_e)} 
\simeq {f_{\tau e}^2 f_\mu^2 \over 32 G_F^2 m_3^4} = 2.6 \times 10^{-7},
\end{equation}
which is well below the experimental upper bound of $1.8 \times 10^{-6} / 
0.1783 = 1.0 \times 10^{-5}$.  Similarly, for $f_{\tau \mu} = 1$, the 
analogous fraction is also $2.6 \times 10^{-7}$ and well below the 
experimental upper bound of $1.9 \times 10^{-6} / 0.1737 = 1.1 \times 
10^{-5}$.  Once produced, the decays of $h_3^0$ are into $s \bar s$, 
$\mu^- \mu^+$, as well as distinct FCNC final states such as $\tau^\pm 
\mu^\mp$, $\tau^\pm e^\mp$, and $s \bar b + b \bar s$.

In conclusion, it has been shown that a supersymmetric extension of the 
standard model with four Higgs doublets has the following desirable 
features.  (1) Only heavy quarks (i.e. $t$, $b$, $c$) and the one heavy 
lepton ($\tau$) are massive under the assumed global U(1) symmetry. 
(2) As the U(1) symmetry is broken softly, the two extra Higgs doublets 
also acquire nonzero (but small) vacuum expectation values, and all the 
light quarks and leptons become massive.  (3) The pattern of the quark 
charged-current mixing matrix is obtained naturally. (4) Small Majorana 
neutrino masses are obtained with three singlet superfields $N_i$ at the 
TeV energy scale.  (5) The two extra Higgs doublets are also at the TeV 
scale with observable decays which are characteristic of this model.
\vskip 0.5in

This work was supported in part by the U.~S.~Department of Energy
under Grant No.~DE-FG03-94ER40837.

\bibliographystyle{unsrt}

\end{document}